\documentclass[aps, prl, twocolumn]{revtex4-1}

\usepackage[pdftex]{graphicx}
\usepackage{amsmath}
\usepackage{amsfonts}
\usepackage{amssymb}
\usepackage{color}
\usepackage{float}
\usepackage{color}
\usepackage{natbib}
\usepackage[utf8]{inputenc}
\usepackage[T1]{fontenc}
\usepackage{siunitx}
\usepackage{textcomp}
\newcommand{\Imag}{\textrm{Im}}

\begin{document}
\title{Nonreciprocal metamaterial obeying time-reversal symmetry}
\author{Siddharth Buddhiraju}
\author{Alex Song}
\author{Georgia T. Papadakis}
\author{Shanhui Fan}
\affiliation{Ginzton Laboratory, Department of Electrical Engineering, Stanford University, Stanford, California 94305, USA}
\date{\today}
\begin{abstract}
We introduce a class of non-Hermitian systems that break electromagnetic reciprocity while preserving time-reversal symmetry, and describe its novel polarization dynamics. We show that this class of systems can be realized using van der Waals heterostructures involving transition-metal dichalcogenides (TMDs). Our work provides a path towards achieving strong optical nonreciprocity and polarization-dependent directional amplification using compact, large-area and magnet-free structures.
\end{abstract}
\maketitle

\textit{Introduction.---} Metamaterials enable us to create systems with effective permittivity and permeability that are not found in nature \citep{pendry2004, alu2005achieving, enghetaMetamaterialsBook, shalaev2007optical, krishnamoorthy2012topological, mahmoud2015all, kivshar2013, xu2013all, monticone2014quest}, opening up new possibilities for manipulating the properties of electromagnetic waves. In general, the allowed forms of the permittivity and permeability tensors are strongly constrained by symmetry considerations. For a nonmagnetic medium with $\mu=1$, the constraints of energy conservation, time-reversal symmetry ($\mathcal{T}$), and reciprocity ($R$) are represented by $\epsilon^\dag=\epsilon$, $\epsilon^*=\epsilon$ and $\epsilon^T=\epsilon$, respectively. Here the exponents $\dag$, $*$ and $T$ denote the Hermitian conjugate, complex conjugate and transpose, respectively. These constraints are related in that any two imply the third. Consequently, in linear, time-invariant media, exactly five classes of dielectric tensors are possible: (1) usual, lossless dielectrics, which satisfy all three conditions, (2) usual, lossy dielectrics, which satisfy only $\epsilon^T=\epsilon$ but not the other two, (3) lossless nonreciprocal media such as magneto-optical materials, which satisfy only $\epsilon^\dag=\epsilon$, (4) nonreciprocal media with time-reversal symmetry, which satisfy only $\epsilon^*=\epsilon$, and (5) media that violate all three constraints. Somewhat surprisingly, among these five classes of materials, there have not been in fact any previous studies of materials or metamaterial systems that belong to class (4), while all other classes of dielectric tensors have been extensively explored. \\

In this Letter, we explore nonreciprocal metamaterial systems that possess time-reversal symmetry. Such systems are, by necessity, non-Hermitian. We first derive the polarization dynamics of such media from their eigevalues and eigenvectors. It will be seen that they possess two distinct phases separated by an exceptional point, akin to $\mathcal{PT}$-symmetric systems. These phases exhibit direction- and polarization- dependent gain and loss and arbitrary polarization conversion for pairs of polarization states, respectively. We show that in the `broken' phase, such media could be used to construct a unique two-way directional amplifier whose direction of amplification depends on the incident polarization. Then, we design a van der Waals heterostructure using transition-metal dichalcogenide (TMD) layers that realizes the proposed dielectric tensor. We conclude with a brief discussion of such nonmagnetic schemes to achieve nonreciprocal behavior. \\

We first outline an important consequence of time-reversal symmetry in such media. Suppose that light of a given polarization propagating through such a medium is amplified by a factor $g$ in one direction. Since the medium obeys $\mathcal{T}$-symmetry, the complex-conjugated polarization must be attenuated by the same factor $g$ when propagating in the reverse direction. As an example, this means that a given linear polarization amplified in forward propagation is attenuated by the same factor in reverse propagation, making the system nonreciprocal. This property distinguishes this class of systems from other non-Hermitian media such as $\mathcal{PT}$-symmetric systems and nonreciprocal directional amplifiers that do not obey $\mathcal{T}$-symmetry. \\

To derive properties of such a $\mathcal{T}$-symmetric non-Hermitian system, we assume that the medium is described by a dielectric tensor of the form
\begin{equation} \label{eq:original}
\bar{\epsilon} = \begin{pmatrix}
\epsilon_{xx} & \epsilon_{xy} & 0\\ \epsilon_{yx} & \epsilon_{yy} & 0 \\ 0 & 0 & \epsilon_{zz}
\end{pmatrix}.
\end{equation}
Under normal incidence, i.e., for propagation along the $\hat{z}$-direction, the relevant part of $\bar{\epsilon}$ is
\begin{equation}
\epsilon_\perp = \begin{pmatrix}
\epsilon_{xx} & \epsilon_{xy}\\ \epsilon_{yx} & \epsilon_{yy} \end{pmatrix}
\label{eq:trs_ep}
\end{equation}
Here, we require all elements of $\epsilon_\perp$ to be real to preserve $\mathcal{T}$-symmetry and $\epsilon_{xy}\neq\epsilon_{yx}$ to break reciprocity. For simplicity, we assume $\epsilon_{xx} = \epsilon_{yy} \equiv \epsilon_r$. Assuming a field solution of the form $\textbf{E}(x,y) e^{-ikz}$, Maxwell's equations in this medium result in an eigenvalue problem of the form
\begin{equation}
k^2\textbf{E}(x,y) = \frac{\omega^2}{c^2}\epsilon_\perp\textbf{E}(x,y).
\end{equation}
The polarization dynamics, i.e., the evolution of the polarization of light as it propagates through this medium, is governed by the eigenvalues and eigenvectors of $\epsilon_\perp$. With $\tau = \epsilon_{yx}/\epsilon_{xy}$, the eigenvalues of $\epsilon_\perp$ are 
\begin{equation}
\epsilon_\pm = \epsilon_r \pm \epsilon_{xy}\sqrt{\tau} \label{eq:eigenvalues},
\end{equation}
and the right eigenvectors are
\begin{equation}
|v^R_\pm\rangle = \frac{1}{\sqrt{2}}\begin{pmatrix}
1 \\ \pm\sqrt{\tau}
\end{pmatrix}. \label{eq:eigenvectors}
\end{equation}
At normal incidence, the two polarizations $|v_\pm^R\rangle$ propagate with wavevectors $k_\pm = k_0\sqrt{\epsilon_\pm}$, where $k_0 = \omega/c$ is the propagation constant in vacuum. The ratio $\tau$ serves as an indicator of broken reciprocity in the medium, since $\epsilon_{yx}\neq \epsilon_{xy}$ corresponds to a nonreciprocal system. The normalization of $|v_\pm^R\rangle$ is set using $\langle v_\pm^L|v_\pm^R\rangle = 1$, where the left eigenvectors $\langle v_\pm^L|$ are defined by the eigenvalue problem $\langle v_\pm^L|\epsilon_\perp = \epsilon_\pm \langle v_\pm^L|$. 

\begin{figure}
\centering
\includegraphics[scale=0.40]{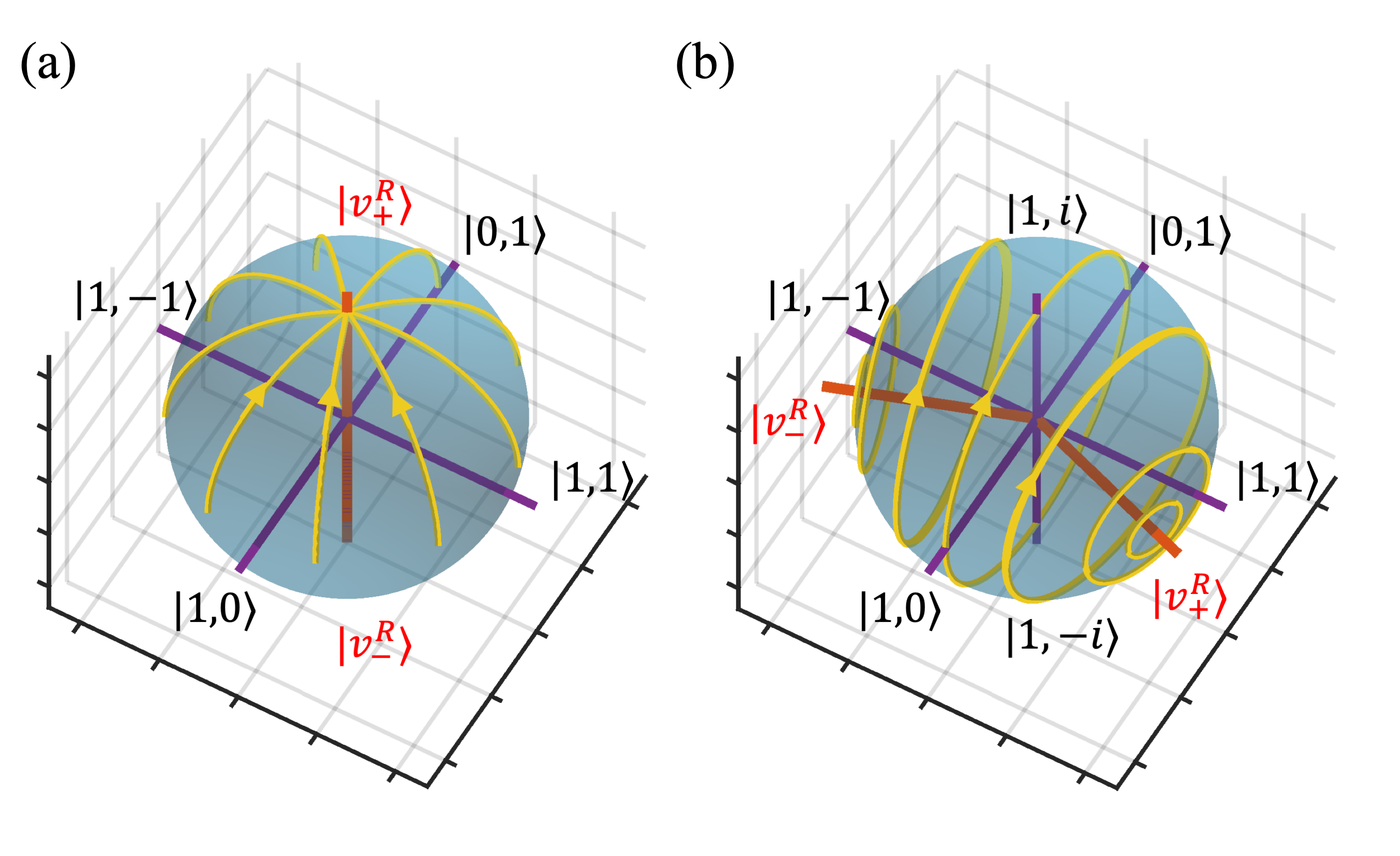}
\caption{Poincare spheres describing the polarization dynamics as light travels through the medium of Eq. \eqref{eq:trs_ep} when (a) $\tau = -1$, and (b) $\tau = +2$, where $\tau = \epsilon_{yx}/\epsilon_{xy}$. When $\tau<0$, all initial polarizations except $|v_-^R\rangle$ amplify to $|v_+^R\rangle$ as they propagate through the medium. For $\tau > 0$, the polarization precesses in circles around the nonorthogonal states $|v_\pm^R\rangle$.}
\label{fig:poincare}
\end{figure}

From Eq. \eqref{eq:eigenvalues}, it is seen that this class of materials possess, in analogy to $\mathcal{PT}$-symmetric systems \citep{cerjan2017, christodoulides2018}, an exact and a broken phase that are separated by an exceptional point. On the other hand, unlike $\mathcal{PT}$-systems, this class of systems preserves $\mathcal{T}$-symmetry in both phases. The exceptional point in this system is given by $\tau=0$, where the eigenvectors coalesce into a single linear polarization. In the broken phase of $\tau<0$, the eigenvectors are two elliptical polarizations that experience equal amounts of gain and loss, respectively. The polarization dynamics in this regime are depicted on the Poincar\' e sphere in Fig. \ref{fig:poincare}(a) for the case of $\tau=-1$. Since the polarization state $|v_+^R\rangle$ experiences gain while $|v_-^R\rangle$ experiences loss, all initial polarization states except $|v_-^R\rangle$ amplify towards $|v_+^R\rangle$ as they propagate through the medium, depicted by the trajectories (yellow lines) on the sphere. On the other hand, in the exact phase of $\tau>0$, the eigenvalues in Eq. \eqref{eq:eigenvalues} are real, corresponding to two linear but nonorthogonal polarizations. Here, the polarization of incident light precesses about the eigenvectors as it travels through the medium. This regime is similar to the exact phase described in Ref. \citep{cerjan2017}, resulting in the ability to transform arbitrary pairs of input polarization states into orthogonal states when the thickness of the medium is chosen appropriately. 

In considering only forward propagation, these systems appear analogous to other non-Hermitian systems. To observe the unique properties of the systems described by Eq. \eqref{eq:trs_ep}, notice that they preserve time-reversal symmetry and break reciprocity. Consequently, when $\tau < 0$, if light of a certain polarization is amplified in forward propagation, its complex conjugate polarization is attenuated by the same factor in reverse propagation, unlike $\mathcal{PT}$-systems. This is clearly seen from Eqs. \eqref{eq:eigenvalues} and \eqref{eq:eigenvectors}, since $\epsilon_+^* = \epsilon_-$ and $|v_+^R\rangle^* = |v_-^R\rangle$ for $\tau < 0$. On the other hand, when $\tau > 0$, the material provides nonreciprocal rotation of polarization. For instance, when the thickness of the medium is chosen to be $d = \pi/2(k_+-k_-)$, $|x\rangle$-polarized light inserted on one end of the medium is rotated and amplified to $\sqrt{\tau}|y\rangle$ on the other end. On the other hand, $|y\rangle$-polarized light inserted back from the other end is rotated and attenuated $\sqrt{\tau}^{-1}|x\rangle$, making it nonreciprocal. Notice that because the factors $\sqrt{\tau}$ and $\sqrt{\tau}^{-1}$ cancel each other, the medium continues to preserve $\mathcal{T}$-symmetry.

\begin{figure}
\centering
\includegraphics[scale=0.5]{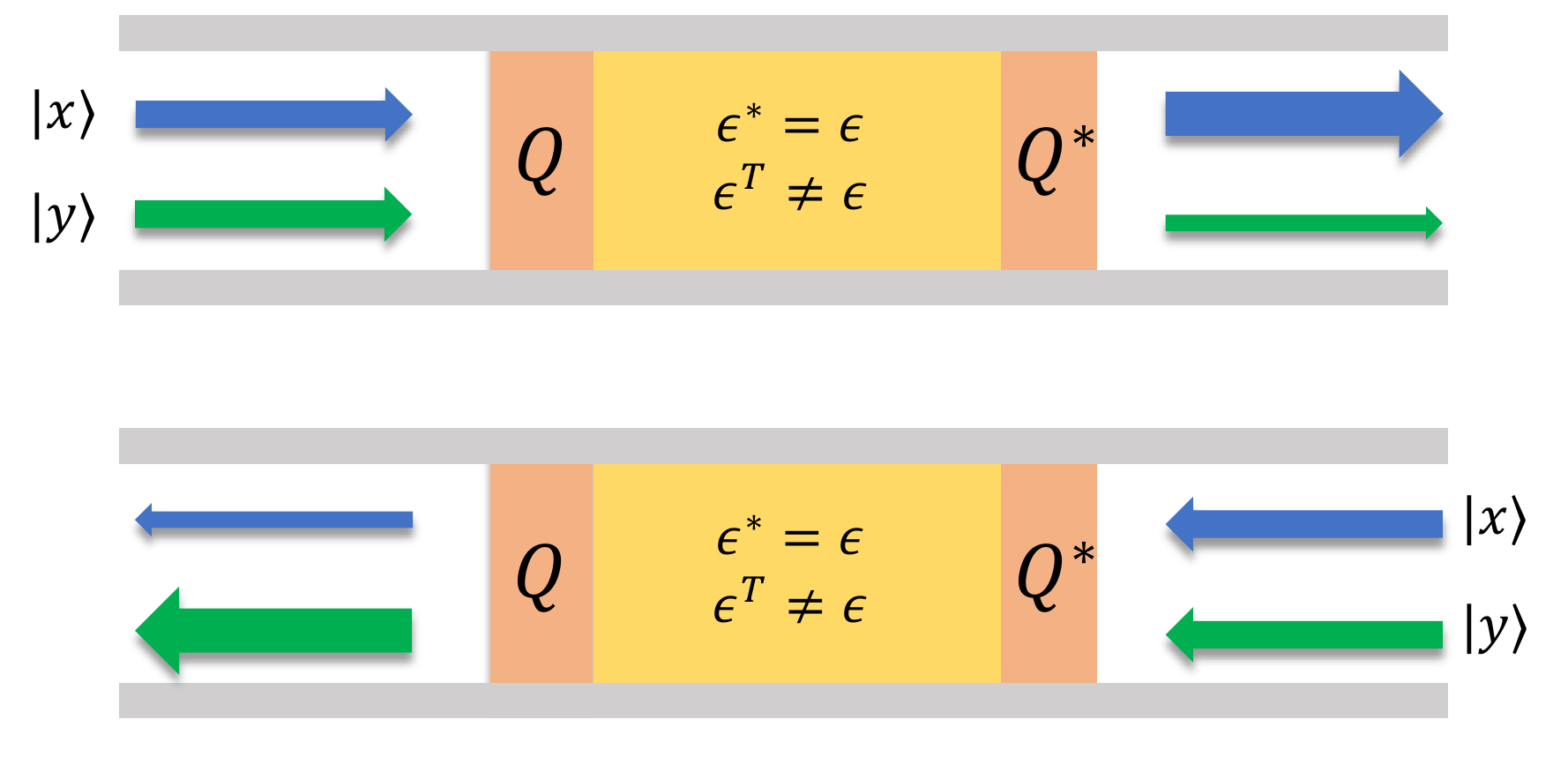}
\caption{Schematic of a communication channel with the time-reversal symmetric metamaterial (yellow) sandwiched between two linear-to-circular polarization converters (orange) with their fast axes rotated by 90\textdegree with respect to each other. In forward propagation (top panel), an $|x\rangle$-polarized signal is converted to $|v_+^R\rangle$, amplified, and rotated back to $|x\rangle$ along the $+z$ direction. On the other hand, a $|y\rangle$ polarized signal is rotated to $|v_-^R\rangle$, attenuated, and rotated to $|y\rangle$ again. In reverse propagation (bottom panel), owing to time-reversal symmetry of the overall structure, an $|x\rangle$-polarized signal is attenuated while a $|y\rangle$-polarized signal is amplified. This results in a unique polarization-dependent two-way directional amplifier.}
\label{fig:apps}
\end{figure}

The presence of time-reversal symmetry in a non-Hermitian system can have interesting consequences. As one example, we show that such media can be used to realize a unique polarization-dependent two-way directional amplifier. In previous works on unidirectional amplification \cite{abdo2013directional, ruesink2016nonreciprocity, kamal2017minimal, koutserimpas2018nonreciprocal, malz2018quantum, song2019direction}, light propagating in one direction is amplified while that in the reverse direction is attenuated. By contrast, the media considered in this work can be used to achieve directional amplification for two orthogonal polarizations in opposite directions. This property can be used in a two-way communication channel, where communication in the forward direction ($+\hat{z}$) is in one polarization, say $|x\rangle$, while that in the backward direction ($-\hat{z}$) is in $|y\rangle$ polarization. Suppose we intend to build a repeater for such a channel, so that $|x\rangle$-polarized light is amplified along $+\hat{z}$ while $|y\rangle$-polarized light is amplified along $-\hat{z}$. Simultaneously, we would like any reflections or noise in $|x\rangle$ polarization to be attenuated in the $-\hat{z}$ direction, and in $|y\rangle$-polarization in the $+\hat{z}$ direction. The schematic for a single unit of such a repeater is shown in Fig. \ref{fig:apps}. This configuration can be achieved using the medium described by Eq. \eqref{eq:trs_ep} with $\epsilon_{xy}=-\epsilon_{yx}$, shown by the purple layer, sandwiched between two linear-to-circular polarization converters (e.g., quarter-wave plates) depicted the by orange layers. Since $\tau= \epsilon_{yx}/\epsilon_{xy}=-1$, the eigenvectors of the medium from Eq. \eqref{eq:eigenvectors} are the two circular polarizations $|v_\pm^R\rangle$, experiencing equal amplification and attenuation, repsectively.  The two linear-to-circular converters have their fast axes orthogonal to each other and can be described using transmission matrices $Q$ and $Q^*$, respectively, where
\begin{equation}
Q = \frac{1}{\sqrt{2}}\begin{pmatrix}
1 & i \\ i & 1
\end{pmatrix}.
\end{equation}
Suppose an $|x\rangle$-polarized signal is inserted into the channel along the $+\hat{z}$ direction. The converter $Q$ rotates $|x\rangle$ to $Q|x\rangle = |v_+^R\rangle$, which is then amplified by the medium. The amplified $|v_+^R\rangle$ signal is then rotated back to an amplified $|x\rangle$ signal by the converter $Q^*$. Notice that because the entire system preserves $\mathcal{T}$-symmetry, any noise or reflection in the $|x\rangle$-polarization traveling along the $-\hat{z}$ direction is attenuated by the medium. Furthermore, since the two polarization converters $Q$ and $Q^*$ have staggered fast axes, it is seen that a $|y\rangle$ signal traveling along the $-\hat{z}$ direction is converted to $Q^*|y\rangle = -i|v_+^R\rangle$, amplified by the same factor, and rotated back to an amplified $|y\rangle$ signal by the converter $Q$. Again, by $\mathcal{T}$-symmetry, any noise or reflection traveling in the $|y\rangle$-polarization along the $+\hat{z}$ direction is attenuated. As a result, this setup acts as a novel two-way directional-amplifier, whose direction of amplification depends on the polarization of the inserted signal. Such a system can be potentially used to achieve enhanced signal-to-noise ration of two-way signal transmission using two orthogonal polarizations. \\ 

As another application, photonic lattices with an imaginary gauge potential \citep{hatano1996localization} have been shown to demonstrate effective unidirectional light transport originating from non-Hermitian delocalization \cite{longhi2015robust, longhi2015non}. Since the $\mathcal{T}$-symmetric media considered here possess direction-dependent amplification and attenuation, they provide a pathway to physically implement an imaginary gauge potential.\\

\textit{Metamaterial design.---} Having discussed the polarization dynamics and potential applications of the class of nonreciprocal media obeying time-reversal symmetry, we now propose a metamaterial design that realizes the general form of Eq. \eqref{eq:trs_ep}. We start with the special case of Eq. \eqref{eq:original} with $\epsilon_{xy}=-\epsilon_{yx}$. In this case, it is seen from Eqs. \eqref{eq:eigenvalues} and \eqref{eq:eigenvectors} that the medium exhibits circular-polarization-dependent gain and loss. To generate this behavior, van der Waals heterostructures, consisting of transition-metal dichalcogenides (TMD) monolayers, are ideal candidates owing to their unique optical properties that are dependent on the circular polarization of incident light. TMD monolayers are gapped Dirac materials with a direct bandgap in the visible frequencies \citep{mak2010}. The broken inversion symmetry in these monolayers results in properties such as the locking of spin- and valley- degrees of freedom, which enable optical addressing of the $K$ and $K'$ valleys with the two circular polarizations \citep{cui2012, mak2016, mak2018}. An interesting consequence of these properties is the valley Hall effect \citep{mceuen2014, schaibley2016}, where upon shining a circularly polarized laser at a TMD monolayer at its bandgap, a population imbalance between the two electron spin states is created, which results in an effective magnetic field owing to the valley-dependent Berry curvature \citep{niu2010}. This effective magnetic field was shown to result in in-plane nonreciprocal transport of plasmon-polaritons \citep{rudner2016, kumar2016}. Moreover, optically pumped TMD monolayers sandwiched in cavities were also shown to achieve lasing \citep{xiang2015, salehzadeh2015}, albeit from both $K$ and $K'$ valleys. In principle, it is possible to achieve circularly polarized lasing by optical pumping with the same circular polarization \citep{mak2016, mak2018} or by injection of a spin-polarized current \citep{rudolph2003}. 


\begin{figure}
\centering
\includegraphics[scale=0.6]{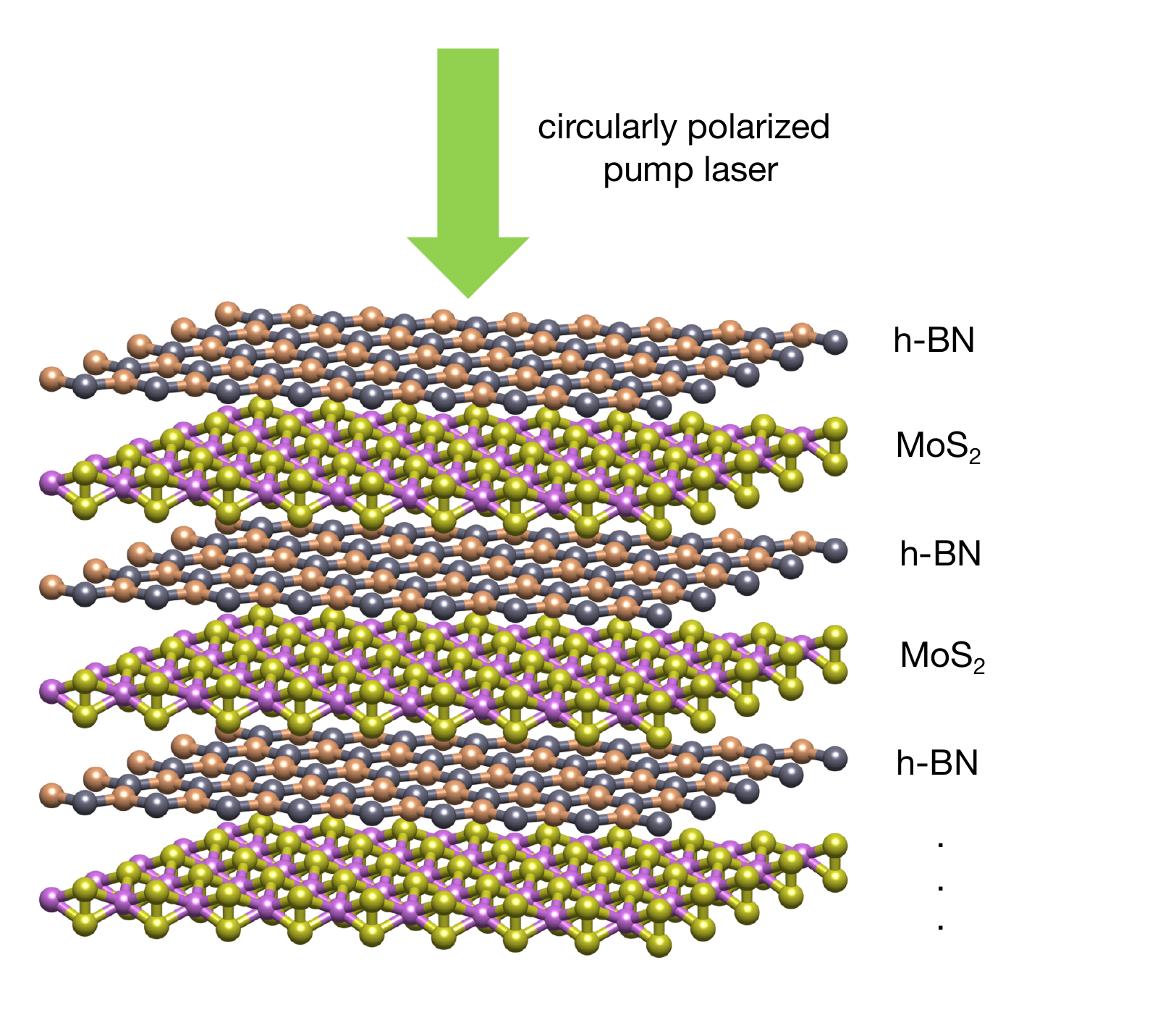}
\caption{A layered van der Waals heterostructure design to realize the dielectric tensor of Eq. \eqref{eq:trs_ep}. The MoS$_2$ monolayers are depicted using pink and yellow spheres. For the special case of $\epsilon_{xy}=-\epsilon_{yx}$, the encapsulation layer is shown by the blue and brown spheres representing hexagonal boron nitride (hBN). The encapsulation ensures that the inversion symmetry of the MoS$_2$ monolayers remains broken. A circularly polarized pump laser shining on the heterostructure provides gain for one of the two valleys of MoS$_2$. For the general form of $\epsilon_{xy}\neq \epsilon_{yx}$, the encapsulation layers can be replaced by any large-bandgap 2D material with in-plane anisotropy, such as SiP or GeP.}
\label{fig:heterostructure}
\end{figure}

Circular polarization-dependent gain in such TMD monolayers provides us a direct means to construct the dielectric tensor of Eq. \eqref{eq:trs_ep}. Consider the structure shown in Fig. \ref{fig:heterostructure}, consisting of molybdenum disulfide (MoS$_2$) monolayers encapsulated between two layers of hexagonal boron nitride (hBN) on each side. hBN is a commonly used encapsulation layer in van der Waals heterostructures \citep{geim2013, dai2015graphene, novoselov2016}, and also serves to maintain the broken inversion symmetry in MoS$_2$ required to utilize its valley-dependent properties. At room temperature, an unpumped MoS$_2$ monolayer is described by an in-plane dielectric constant of $\epsilon_\perp = 22.773 - 11.177i$, a layer thickness of $6.15$ \AA  \citep{heinz2014}, and a longitudinal dielectric constant of $\epsilon_{||} = 6.4$ \citep{laturia2018}. For hBN,  we use dielectric constants $\epsilon_{\perp} = 4.855$ and $\epsilon_{||} = 2.949$ \citep{bozhao2017} with a layer thickness of 3.17 \AA \citep{laturia2018}. In principle, one of the two valleys of MoS$_2$ can be completely inverted by pumping using a circularly polarized laser, resulting in very strong optical gain as well as nonreciprocity. However, since such strong gain in TMD monolayers is yet to be demonstrated \citep{chernikov2015}, we assume that the MoS$_2$ monolayer is first uniformly pumped to $\epsilon_\perp = 22.773 - 1i$, whereafter a circularly polarized laser at a frequency slightly above the MoS$_2$ bandgap of 1.9 eV induces a gain of $\Imag(\epsilon_\perp) = +2i$ for one of the two valleys. In the basis of the two circular polarizations, the in-plane dielectric tensor of the pumped MoS$_2$ monolayer is
\begin{equation}
\epsilon^{cc}_\perp = \begin{pmatrix}
22.773 + 1i & 0 \\ 0 & 22.773 - 1i
\end{pmatrix}.
\end{equation}
By a unitary transform, this dielectric tensor can be converted to the $x$-$y$ basis:
\begin{equation}
\epsilon_\perp = \begin{pmatrix}
\epsilon_r & \epsilon_{xy} \\ \epsilon_{yx} & \epsilon_r
\end{pmatrix}, \ \epsilon_r = 22.773, \ \epsilon_{xy}=-\epsilon_{yx}=-1,
\label{eq:main}
\end{equation}
which is in the form expected by Eq. \eqref{eq:trs_ep}. In order to derive the dielectric tensor of the heterostructure of Fig. \ref{fig:heterostructure}, we employ effective medium theory \cite{agranovich1985notes}. The effective medium approximation applies to the deep subwavelength limit, which is valid here since the unit cell consisting of an MoS$_2$ monolayer sandwitched between two layers of hBN is significantly thinner than the operating wavelength of approximately 650 nm. In this approximation, the effective dielectric tensor of the van der Waals heterostructure of Fig. \ref{fig:heterostructure} is given by
\begin{align}
\epsilon_{\perp} &= \begin{pmatrix}
10.7071 & -0.3266 \\ 0.3266 & 10.7071
\end{pmatrix}, \label{eq:effectivemedium}\\
\epsilon_{||} &= 3.5794.
\end{align}
A comparison of these values with parameter retrieval \citep{smith2005} from an exact transfer matrix calculation resulted in a difference of less than 0.1\%, confirming the predictions of effective medium theory in this deep subwavelength limit. Thus far, we have constructed a metamaterial that realizes Eq. \eqref{eq:trs_ep} with $\epsilon_{xy}=-\epsilon_{yx}$. \\

In order to realize the general form of Eq. \eqref{eq:trs_ep}, we insert an anisotropic 2D material layer into the heterostructure. Several choices for such in-plane anisotropy exist \citep{BP1, BP2, SiP, GeP, GeAs2}. Since we would like the anisotropic layer to be nearly lossless at near the MoS$_2$ bandgap of 1.9 eV, SiP \citep{SiP} or GeP \citep{GeP}, which have direct bandgaps at around 2.5 eV, are good candidates. In general, lossless materials possessing an in-plane anisotropy can be represented by
\begin{equation} 
\epsilon_\perp = \begin{pmatrix}
\epsilon_r' & \delta \\ \delta & \epsilon_r'
\end{pmatrix}, \ \ \epsilon_r', \delta \in \mathbb{R}
\label{eq:birefringent}
\end{equation}
where $\delta$ is the degree of anisotropy. Consider an effective medium constructed using the time-reversal preserving metamaterial of Eq. \eqref{eq:main} and the anisotropic layer of Eq. \eqref{eq:birefringent}. In the resulting effective medium tensor, the elements $\epsilon_{xy}$ of Eq. \eqref{eq:main} and $\delta$ of Eq. \eqref{eq:birefringent} add up in one of the off-diagonal components but subtract in the other. This allows us to achieve two real but unequal off-diagonal components. Furthermore, by designing their amplitudes appropriately by varying the gain in the MoS$_2$ monolayer or varying the number of anisotropic monolayers, it is possible to have both off-diagonal components in the effective medium to have the same sign. Therefore, the general form of Eq. \eqref{eq:trs_ep} can be achieved by a van der Waals heterostructure consisting of TMD monolayers pumped with circularly polarized light, in-plane anisotropic media such as SiP or GeP, and/or appropriate encapsulation layers such as hBN.

\textit{Discussion.---} In this Letter, we explore a new class of non-Hermitian, nonreciprocal metamaterials that preserve time-reversal symmetry. From their polarization dynamics, we highlight their unique properties such as polarization- and direction- dependent gain and loss that enables two-way signal amplification while simultaneously attenuating back-reflections in both directions. We further propose a design for such metamaterials in the form of van der Waals heterostructures comprising 2D materials such as transition metal dichalcogenides (TMDs) and anistropic monolayers. More generally, nonreciprocal designs based on 2D materials, such as the ones proposed in this Letter, allow for large-area nonreciprocity, providing a potential advantage over nonreciprocity arising from time-modulated systems \citep{zongfuTimeModulation2009, kejieTimeModulation2012, aluTimeModulation2017} that are constrained in their cross-sectional area. Furthermore, the strength of the nonreciprocity depends only on the ability to pump TMD monolayers using circularly polarized lasers, unlike magneto-optical systems where substantial external magnetic fields may be required to observe significant nonreciprocity. \\

\begin{acknowledgements}
This work is supported by the U. S. National Science Foundation (Grant No. CBET-1641069), and  by the Department of Defense Joint Directed Energy Transition Office (DE-JTO) under Grant No. N00014-17-1-2557. S.B. acknowledges the support of a Stanford Graduate Fellowship. G.T.P. acknowledges financial support from the TomKat Center for Sustainable Energy at Stanford University. S.B. thanks Wei Li, Avik Dutt, Ian Williamson, and Elyse Barr\' e for valuable discussions. 
\end{acknowledgements}

\begin{thebibliography}{49}
\expandafter\ifx\csname natexlab\endcsname\relax\def\natexlab#1{#1}\fi
\expandafter\ifx\csname bibnamefont\endcsname\relax
  \def\bibnamefont#1{#1}\fi
\expandafter\ifx\csname bibfnamefont\endcsname\relax
  \def\bibfnamefont#1{#1}\fi
\expandafter\ifx\csname citenamefont\endcsname\relax
  \def\citenamefont#1{#1}\fi
\expandafter\ifx\csname url\endcsname\relax
  \def\url#1{\texttt{#1}}\fi
\expandafter\ifx\csname urlprefix\endcsname\relax\def\urlprefix{URL }\fi
\providecommand{\bibinfo}[2]{#2}
\providecommand{\eprint}[2][]{\url{#2}}

\bibitem[{\citenamefont{Smith et~al.}(2004)\citenamefont{Smith, Pendry, and
  Wiltshire}}]{pendry2004}
\bibinfo{author}{\bibfnamefont{D.~R.} \bibnamefont{Smith}},
  \bibinfo{author}{\bibfnamefont{J.~B.} \bibnamefont{Pendry}},
  \bibnamefont{and} \bibinfo{author}{\bibfnamefont{M.~C.}
  \bibnamefont{Wiltshire}}, \bibinfo{journal}{Science}
  \textbf{\bibinfo{volume}{305}}, \bibinfo{pages}{788} (\bibinfo{year}{2004}).

\bibitem[{\citenamefont{Al{\`u} and Engheta}(2005)}]{alu2005achieving}
\bibinfo{author}{\bibfnamefont{A.}~\bibnamefont{Al{\`u}}} \bibnamefont{and}
  \bibinfo{author}{\bibfnamefont{N.}~\bibnamefont{Engheta}},
  \bibinfo{journal}{Physical Review E} \textbf{\bibinfo{volume}{72}},
  \bibinfo{pages}{016623} (\bibinfo{year}{2005}).

\bibitem[{\citenamefont{Engheta and
  Ziolkowski}(2006)}]{enghetaMetamaterialsBook}
\bibinfo{author}{\bibfnamefont{N.}~\bibnamefont{Engheta}} \bibnamefont{and}
  \bibinfo{author}{\bibfnamefont{R.~W.} \bibnamefont{Ziolkowski}},
  \emph{\bibinfo{title}{Metamaterials: physics and engineering explorations}}
  (\bibinfo{publisher}{John Wiley \& Sons}, \bibinfo{year}{2006}).

\bibitem[{\citenamefont{Shalaev}(2007)}]{shalaev2007optical}
\bibinfo{author}{\bibfnamefont{V.~M.} \bibnamefont{Shalaev}},
  \bibinfo{journal}{Nature Photonics} \textbf{\bibinfo{volume}{1}},
  \bibinfo{pages}{41} (\bibinfo{year}{2007}).

\bibitem[{\citenamefont{Krishnamoorthy
  et~al.}(2012)\citenamefont{Krishnamoorthy, Jacob, Narimanov, Kretzschmar, and
  Menon}}]{krishnamoorthy2012topological}
\bibinfo{author}{\bibfnamefont{H.~N.} \bibnamefont{Krishnamoorthy}},
  \bibinfo{author}{\bibfnamefont{Z.}~\bibnamefont{Jacob}},
  \bibinfo{author}{\bibfnamefont{E.}~\bibnamefont{Narimanov}},
  \bibinfo{author}{\bibfnamefont{I.}~\bibnamefont{Kretzschmar}},
  \bibnamefont{and} \bibinfo{author}{\bibfnamefont{V.~M.} \bibnamefont{Menon}},
  \bibinfo{journal}{Science} \textbf{\bibinfo{volume}{336}},
  \bibinfo{pages}{205} (\bibinfo{year}{2012}).

\bibitem[{\citenamefont{Mahmoud et~al.}(2015)\citenamefont{Mahmoud, Davoyan,
  and Engheta}}]{mahmoud2015all}
\bibinfo{author}{\bibfnamefont{A.~M.} \bibnamefont{Mahmoud}},
  \bibinfo{author}{\bibfnamefont{A.~R.} \bibnamefont{Davoyan}},
  \bibnamefont{and} \bibinfo{author}{\bibfnamefont{N.}~\bibnamefont{Engheta}},
  \bibinfo{journal}{Nature Communications} \textbf{\bibinfo{volume}{6}},
  \bibinfo{pages}{8359} (\bibinfo{year}{2015}).

\bibitem[{\citenamefont{Poddubny et~al.}(2013)\citenamefont{Poddubny, Iorsh,
  Belov, and Kivshar}}]{kivshar2013}
\bibinfo{author}{\bibfnamefont{A.}~\bibnamefont{Poddubny}},
  \bibinfo{author}{\bibfnamefont{I.}~\bibnamefont{Iorsh}},
  \bibinfo{author}{\bibfnamefont{P.}~\bibnamefont{Belov}}, \bibnamefont{and}
  \bibinfo{author}{\bibfnamefont{Y.}~\bibnamefont{Kivshar}},
  \bibinfo{journal}{Nature Photonics} \textbf{\bibinfo{volume}{7}},
  \bibinfo{pages}{948} (\bibinfo{year}{2013}).

\bibitem[{\citenamefont{Xu et~al.}(2013)\citenamefont{Xu, Agrawal, Abashin,
  Chau, and Lezec}}]{xu2013all}
\bibinfo{author}{\bibfnamefont{T.}~\bibnamefont{Xu}},
  \bibinfo{author}{\bibfnamefont{A.}~\bibnamefont{Agrawal}},
  \bibinfo{author}{\bibfnamefont{M.}~\bibnamefont{Abashin}},
  \bibinfo{author}{\bibfnamefont{K.~J.} \bibnamefont{Chau}}, \bibnamefont{and}
  \bibinfo{author}{\bibfnamefont{H.~J.} \bibnamefont{Lezec}},
  \bibinfo{journal}{Nature} \textbf{\bibinfo{volume}{497}},
  \bibinfo{pages}{470} (\bibinfo{year}{2013}).

\bibitem[{\citenamefont{Monticone and Alu}(2014)}]{monticone2014quest}
\bibinfo{author}{\bibfnamefont{F.}~\bibnamefont{Monticone}} \bibnamefont{and}
  \bibinfo{author}{\bibfnamefont{A.}~\bibnamefont{Alu}},
  \bibinfo{journal}{Journal of Materials Chemistry C}
  \textbf{\bibinfo{volume}{2}}, \bibinfo{pages}{9059} (\bibinfo{year}{2014}).

\bibitem[{\citenamefont{Cerjan and Fan}(2017)}]{cerjan2017}
\bibinfo{author}{\bibfnamefont{A.}~\bibnamefont{Cerjan}} \bibnamefont{and}
  \bibinfo{author}{\bibfnamefont{S.}~\bibnamefont{Fan}},
  \bibinfo{journal}{Physical Review Letters} \textbf{\bibinfo{volume}{118}},
  \bibinfo{pages}{253902} (\bibinfo{year}{2017}).

\bibitem[{\citenamefont{El-Ganainy et~al.}(2018)\citenamefont{El-Ganainy,
  Makris, Khajavikhan, Musslimani, Rotter, and
  Christodoulides}}]{christodoulides2018}
\bibinfo{author}{\bibfnamefont{R.}~\bibnamefont{El-Ganainy}},
  \bibinfo{author}{\bibfnamefont{K.~G.} \bibnamefont{Makris}},
  \bibinfo{author}{\bibfnamefont{M.}~\bibnamefont{Khajavikhan}},
  \bibinfo{author}{\bibfnamefont{Z.~H.} \bibnamefont{Musslimani}},
  \bibinfo{author}{\bibfnamefont{S.}~\bibnamefont{Rotter}}, \bibnamefont{and}
  \bibinfo{author}{\bibfnamefont{D.~N.} \bibnamefont{Christodoulides}},
  \bibinfo{journal}{Nature Physics} \textbf{\bibinfo{volume}{14}},
  \bibinfo{pages}{11} (\bibinfo{year}{2018}).

\bibitem[{\citenamefont{Abdo et~al.}(2013)\citenamefont{Abdo, Sliwa, Frunzio,
  and Devoret}}]{abdo2013directional}
\bibinfo{author}{\bibfnamefont{B.}~\bibnamefont{Abdo}},
  \bibinfo{author}{\bibfnamefont{K.}~\bibnamefont{Sliwa}},
  \bibinfo{author}{\bibfnamefont{L.}~\bibnamefont{Frunzio}}, \bibnamefont{and}
  \bibinfo{author}{\bibfnamefont{M.}~\bibnamefont{Devoret}},
  \bibinfo{journal}{Physical Review X} \textbf{\bibinfo{volume}{3}},
  \bibinfo{pages}{031001} (\bibinfo{year}{2013}).

\bibitem[{\citenamefont{Ruesink et~al.}(2016)\citenamefont{Ruesink, Miri, Alu,
  and Verhagen}}]{ruesink2016nonreciprocity}
\bibinfo{author}{\bibfnamefont{F.}~\bibnamefont{Ruesink}},
  \bibinfo{author}{\bibfnamefont{M.-A.} \bibnamefont{Miri}},
  \bibinfo{author}{\bibfnamefont{A.}~\bibnamefont{Alu}}, \bibnamefont{and}
  \bibinfo{author}{\bibfnamefont{E.}~\bibnamefont{Verhagen}},
  \bibinfo{journal}{Nature Communications} \textbf{\bibinfo{volume}{7}},
  \bibinfo{pages}{13662} (\bibinfo{year}{2016}).

\bibitem[{\citenamefont{Kamal and Metelmann}(2017)}]{kamal2017minimal}
\bibinfo{author}{\bibfnamefont{A.}~\bibnamefont{Kamal}} \bibnamefont{and}
  \bibinfo{author}{\bibfnamefont{A.}~\bibnamefont{Metelmann}},
  \bibinfo{journal}{Physical Review Applied} \textbf{\bibinfo{volume}{7}},
  \bibinfo{pages}{034031} (\bibinfo{year}{2017}).

\bibitem[{\citenamefont{Koutserimpas and
  Fleury}(2018)}]{koutserimpas2018nonreciprocal}
\bibinfo{author}{\bibfnamefont{T.~T.} \bibnamefont{Koutserimpas}}
  \bibnamefont{and} \bibinfo{author}{\bibfnamefont{R.}~\bibnamefont{Fleury}},
  \bibinfo{journal}{Physical Review Letters} \textbf{\bibinfo{volume}{120}},
  \bibinfo{pages}{087401} (\bibinfo{year}{2018}).

\bibitem[{\citenamefont{Malz et~al.}(2018)\citenamefont{Malz, T{\'o}th,
  Bernier, Feofanov, Kippenberg, and Nunnenkamp}}]{malz2018quantum}
\bibinfo{author}{\bibfnamefont{D.}~\bibnamefont{Malz}},
  \bibinfo{author}{\bibfnamefont{L.~D.} \bibnamefont{T{\'o}th}},
  \bibinfo{author}{\bibfnamefont{N.~R.} \bibnamefont{Bernier}},
  \bibinfo{author}{\bibfnamefont{A.~K.} \bibnamefont{Feofanov}},
  \bibinfo{author}{\bibfnamefont{T.~J.} \bibnamefont{Kippenberg}},
  \bibnamefont{and}
  \bibinfo{author}{\bibfnamefont{A.}~\bibnamefont{Nunnenkamp}},
  \bibinfo{journal}{Physical Review Letters} \textbf{\bibinfo{volume}{120}},
  \bibinfo{pages}{023601} (\bibinfo{year}{2018}).

\bibitem[{\citenamefont{Song et~al.}(2019)\citenamefont{Song, Shi, Lin, and
  Fan}}]{song2019direction}
\bibinfo{author}{\bibfnamefont{A.~Y.} \bibnamefont{Song}},
  \bibinfo{author}{\bibfnamefont{Y.}~\bibnamefont{Shi}},
  \bibinfo{author}{\bibfnamefont{Q.}~\bibnamefont{Lin}}, \bibnamefont{and}
  \bibinfo{author}{\bibfnamefont{S.}~\bibnamefont{Fan}},
  \bibinfo{journal}{Physical Review A} \textbf{\bibinfo{volume}{99}},
  \bibinfo{pages}{013824} (\bibinfo{year}{2019}).

\bibitem[{\citenamefont{Hatano and Nelson}(1996)}]{hatano1996localization}
\bibinfo{author}{\bibfnamefont{N.}~\bibnamefont{Hatano}} \bibnamefont{and}
  \bibinfo{author}{\bibfnamefont{D.~R.} \bibnamefont{Nelson}},
  \bibinfo{journal}{Physical Review Letters} \textbf{\bibinfo{volume}{77}},
  \bibinfo{pages}{570} (\bibinfo{year}{1996}).

\bibitem[{\citenamefont{Longhi et~al.}(2015{\natexlab{a}})\citenamefont{Longhi,
  Gatti, and Della~Valle}}]{longhi2015robust}
\bibinfo{author}{\bibfnamefont{S.}~\bibnamefont{Longhi}},
  \bibinfo{author}{\bibfnamefont{D.}~\bibnamefont{Gatti}}, \bibnamefont{and}
  \bibinfo{author}{\bibfnamefont{G.}~\bibnamefont{Della~Valle}},
  \bibinfo{journal}{Scientific Reports} \textbf{\bibinfo{volume}{5}},
  \bibinfo{pages}{13376} (\bibinfo{year}{2015}{\natexlab{a}}).

\bibitem[{\citenamefont{Longhi et~al.}(2015{\natexlab{b}})\citenamefont{Longhi,
  Gatti, and Della~Valle}}]{longhi2015non}
\bibinfo{author}{\bibfnamefont{S.}~\bibnamefont{Longhi}},
  \bibinfo{author}{\bibfnamefont{D.}~\bibnamefont{Gatti}}, \bibnamefont{and}
  \bibinfo{author}{\bibfnamefont{G.}~\bibnamefont{Della~Valle}},
  \bibinfo{journal}{Physical Review B} \textbf{\bibinfo{volume}{92}},
  \bibinfo{pages}{094204} (\bibinfo{year}{2015}{\natexlab{b}}).

\bibitem[{\citenamefont{Mak et~al.}(2010)\citenamefont{Mak, Lee, Hone, Shan,
  and Heinz}}]{mak2010}
\bibinfo{author}{\bibfnamefont{K.~F.} \bibnamefont{Mak}},
  \bibinfo{author}{\bibfnamefont{C.}~\bibnamefont{Lee}},
  \bibinfo{author}{\bibfnamefont{J.}~\bibnamefont{Hone}},
  \bibinfo{author}{\bibfnamefont{J.}~\bibnamefont{Shan}}, \bibnamefont{and}
  \bibinfo{author}{\bibfnamefont{T.~F.} \bibnamefont{Heinz}},
  \bibinfo{journal}{Physical Review Letters} \textbf{\bibinfo{volume}{105}},
  \bibinfo{pages}{136805} (\bibinfo{year}{2010}).

\bibitem[{\citenamefont{Zeng et~al.}(2012)\citenamefont{Zeng, Dai, Yao, Xiao,
  and Cui}}]{cui2012}
\bibinfo{author}{\bibfnamefont{H.}~\bibnamefont{Zeng}},
  \bibinfo{author}{\bibfnamefont{J.}~\bibnamefont{Dai}},
  \bibinfo{author}{\bibfnamefont{W.}~\bibnamefont{Yao}},
  \bibinfo{author}{\bibfnamefont{D.}~\bibnamefont{Xiao}}, \bibnamefont{and}
  \bibinfo{author}{\bibfnamefont{X.}~\bibnamefont{Cui}},
  \bibinfo{journal}{Nature Nanotechnology} \textbf{\bibinfo{volume}{7}},
  \bibinfo{pages}{490} (\bibinfo{year}{2012}).

\bibitem[{\citenamefont{Mak and Shan}(2016)}]{mak2016}
\bibinfo{author}{\bibfnamefont{K.~F.} \bibnamefont{Mak}} \bibnamefont{and}
  \bibinfo{author}{\bibfnamefont{J.}~\bibnamefont{Shan}},
  \bibinfo{journal}{Nature Photonics} \textbf{\bibinfo{volume}{10}},
  \bibinfo{pages}{216} (\bibinfo{year}{2016}).

\bibitem[{\citenamefont{Mak et~al.}(2018)\citenamefont{Mak, Xiao, and
  Shan}}]{mak2018}
\bibinfo{author}{\bibfnamefont{K.~F.} \bibnamefont{Mak}},
  \bibinfo{author}{\bibfnamefont{D.}~\bibnamefont{Xiao}}, \bibnamefont{and}
  \bibinfo{author}{\bibfnamefont{J.}~\bibnamefont{Shan}},
  \bibinfo{journal}{Nature Photonics} \textbf{\bibinfo{volume}{12}},
  \bibinfo{pages}{451} (\bibinfo{year}{2018}).

\bibitem[{\citenamefont{Mak et~al.}(2014)\citenamefont{Mak, McGill, Park, and
  McEuen}}]{mceuen2014}
\bibinfo{author}{\bibfnamefont{K.~F.} \bibnamefont{Mak}},
  \bibinfo{author}{\bibfnamefont{K.~L.} \bibnamefont{McGill}},
  \bibinfo{author}{\bibfnamefont{J.}~\bibnamefont{Park}}, \bibnamefont{and}
  \bibinfo{author}{\bibfnamefont{P.~L.} \bibnamefont{McEuen}},
  \bibinfo{journal}{Science} \textbf{\bibinfo{volume}{344}},
  \bibinfo{pages}{1489} (\bibinfo{year}{2014}).

\bibitem[{\citenamefont{Schaibley et~al.}(2016)\citenamefont{Schaibley, Yu,
  Clark, Rivera, Ross, Seyler, Yao, and Xu}}]{schaibley2016}
\bibinfo{author}{\bibfnamefont{J.~R.} \bibnamefont{Schaibley}},
  \bibinfo{author}{\bibfnamefont{H.}~\bibnamefont{Yu}},
  \bibinfo{author}{\bibfnamefont{G.}~\bibnamefont{Clark}},
  \bibinfo{author}{\bibfnamefont{P.}~\bibnamefont{Rivera}},
  \bibinfo{author}{\bibfnamefont{J.~S.} \bibnamefont{Ross}},
  \bibinfo{author}{\bibfnamefont{K.~L.} \bibnamefont{Seyler}},
  \bibinfo{author}{\bibfnamefont{W.}~\bibnamefont{Yao}}, \bibnamefont{and}
  \bibinfo{author}{\bibfnamefont{X.}~\bibnamefont{Xu}},
  \bibinfo{journal}{Nature Reviews Materials} \textbf{\bibinfo{volume}{1}},
  \bibinfo{pages}{16055} (\bibinfo{year}{2016}).

\bibitem[{\citenamefont{Xiao et~al.}(2010)\citenamefont{Xiao, Chang, and
  Niu}}]{niu2010}
\bibinfo{author}{\bibfnamefont{D.}~\bibnamefont{Xiao}},
  \bibinfo{author}{\bibfnamefont{M.-C.} \bibnamefont{Chang}}, \bibnamefont{and}
  \bibinfo{author}{\bibfnamefont{Q.}~\bibnamefont{Niu}},
  \bibinfo{journal}{Reviews of Modern Physics} \textbf{\bibinfo{volume}{82}},
  \bibinfo{pages}{1959} (\bibinfo{year}{2010}).

\bibitem[{\citenamefont{Song and Rudner}(2016)}]{rudner2016}
\bibinfo{author}{\bibfnamefont{J.~C.} \bibnamefont{Song}} \bibnamefont{and}
  \bibinfo{author}{\bibfnamefont{M.~S.} \bibnamefont{Rudner}},
  \bibinfo{journal}{Proceedings of the National Academy of Sciences}
  \textbf{\bibinfo{volume}{113}}, \bibinfo{pages}{4658} (\bibinfo{year}{2016}).

\bibitem[{\citenamefont{Kumar et~al.}(2016)\citenamefont{Kumar, Nemilentsau,
  Fung, Hanson, Fang, and Low}}]{kumar2016}
\bibinfo{author}{\bibfnamefont{A.}~\bibnamefont{Kumar}},
  \bibinfo{author}{\bibfnamefont{A.}~\bibnamefont{Nemilentsau}},
  \bibinfo{author}{\bibfnamefont{K.~H.} \bibnamefont{Fung}},
  \bibinfo{author}{\bibfnamefont{G.}~\bibnamefont{Hanson}},
  \bibinfo{author}{\bibfnamefont{N.~X.} \bibnamefont{Fang}}, \bibnamefont{and}
  \bibinfo{author}{\bibfnamefont{T.}~\bibnamefont{Low}},
  \bibinfo{journal}{Physical Review B} \textbf{\bibinfo{volume}{93}},
  \bibinfo{pages}{041413} (\bibinfo{year}{2016}).

\bibitem[{\citenamefont{Ye et~al.}(2015)\citenamefont{Ye, Wong, Lu, Ni, Zhu,
  Chen, Wang, and Zhang}}]{xiang2015}
\bibinfo{author}{\bibfnamefont{Y.}~\bibnamefont{Ye}},
  \bibinfo{author}{\bibfnamefont{Z.~J.} \bibnamefont{Wong}},
  \bibinfo{author}{\bibfnamefont{X.}~\bibnamefont{Lu}},
  \bibinfo{author}{\bibfnamefont{X.}~\bibnamefont{Ni}},
  \bibinfo{author}{\bibfnamefont{H.}~\bibnamefont{Zhu}},
  \bibinfo{author}{\bibfnamefont{X.}~\bibnamefont{Chen}},
  \bibinfo{author}{\bibfnamefont{Y.}~\bibnamefont{Wang}}, \bibnamefont{and}
  \bibinfo{author}{\bibfnamefont{X.}~\bibnamefont{Zhang}},
  \bibinfo{journal}{Nature Photonics} \textbf{\bibinfo{volume}{9}},
  \bibinfo{pages}{733} (\bibinfo{year}{2015}).

\bibitem[{\citenamefont{Salehzadeh et~al.}(2015)\citenamefont{Salehzadeh,
  Djavid, Tran, Shih, and Mi}}]{salehzadeh2015}
\bibinfo{author}{\bibfnamefont{O.}~\bibnamefont{Salehzadeh}},
  \bibinfo{author}{\bibfnamefont{M.}~\bibnamefont{Djavid}},
  \bibinfo{author}{\bibfnamefont{N.~H.} \bibnamefont{Tran}},
  \bibinfo{author}{\bibfnamefont{I.}~\bibnamefont{Shih}}, \bibnamefont{and}
  \bibinfo{author}{\bibfnamefont{Z.}~\bibnamefont{Mi}}, \bibinfo{journal}{Nano
  Letters} \textbf{\bibinfo{volume}{15}}, \bibinfo{pages}{5302}
  (\bibinfo{year}{2015}).

\bibitem[{\citenamefont{Rudolph et~al.}(2003)\citenamefont{Rudolph, H{\"a}gele,
  Gibbs, Khitrova, and Oestreich}}]{rudolph2003}
\bibinfo{author}{\bibfnamefont{J.}~\bibnamefont{Rudolph}},
  \bibinfo{author}{\bibfnamefont{D.}~\bibnamefont{H{\"a}gele}},
  \bibinfo{author}{\bibfnamefont{H.}~\bibnamefont{Gibbs}},
  \bibinfo{author}{\bibfnamefont{G.}~\bibnamefont{Khitrova}}, \bibnamefont{and}
  \bibinfo{author}{\bibfnamefont{M.}~\bibnamefont{Oestreich}},
  \bibinfo{journal}{Applied Physics Letters} \textbf{\bibinfo{volume}{82}},
  \bibinfo{pages}{4516} (\bibinfo{year}{2003}).

\bibitem[{\citenamefont{Geim and Grigorieva}(2013)}]{geim2013}
\bibinfo{author}{\bibfnamefont{A.~K.} \bibnamefont{Geim}} \bibnamefont{and}
  \bibinfo{author}{\bibfnamefont{I.~V.} \bibnamefont{Grigorieva}},
  \bibinfo{journal}{Nature} \textbf{\bibinfo{volume}{499}},
  \bibinfo{pages}{419} (\bibinfo{year}{2013}).

\bibitem[{\citenamefont{Dai et~al.}(2015)\citenamefont{Dai, Ma, Liu, Andersen,
  Fei, Goldflam, Wagner, Watanabe, Taniguchi, Thiemens
  et~al.}}]{dai2015graphene}
\bibinfo{author}{\bibfnamefont{S.}~\bibnamefont{Dai}},
  \bibinfo{author}{\bibfnamefont{Q.}~\bibnamefont{Ma}},
  \bibinfo{author}{\bibfnamefont{M.}~\bibnamefont{Liu}},
  \bibinfo{author}{\bibfnamefont{T.}~\bibnamefont{Andersen}},
  \bibinfo{author}{\bibfnamefont{Z.}~\bibnamefont{Fei}},
  \bibinfo{author}{\bibfnamefont{M.}~\bibnamefont{Goldflam}},
  \bibinfo{author}{\bibfnamefont{M.}~\bibnamefont{Wagner}},
  \bibinfo{author}{\bibfnamefont{K.}~\bibnamefont{Watanabe}},
  \bibinfo{author}{\bibfnamefont{T.}~\bibnamefont{Taniguchi}},
  \bibinfo{author}{\bibfnamefont{M.}~\bibnamefont{Thiemens}},
  \bibnamefont{et~al.}, \bibinfo{journal}{Nature Nanotechnology}
  \textbf{\bibinfo{volume}{10}}, \bibinfo{pages}{682} (\bibinfo{year}{2015}).

\bibitem[{\citenamefont{Novoselov et~al.}(2016)\citenamefont{Novoselov,
  Mishchenko, Carvalho, and Neto}}]{novoselov2016}
\bibinfo{author}{\bibfnamefont{K.}~\bibnamefont{Novoselov}},
  \bibinfo{author}{\bibfnamefont{A.}~\bibnamefont{Mishchenko}},
  \bibinfo{author}{\bibfnamefont{A.}~\bibnamefont{Carvalho}}, \bibnamefont{and}
  \bibinfo{author}{\bibfnamefont{A.~C.} \bibnamefont{Neto}},
  \bibinfo{journal}{Science} \textbf{\bibinfo{volume}{353}},
  \bibinfo{pages}{aac9439} (\bibinfo{year}{2016}).

\bibitem[{\citenamefont{Li et~al.}(2014)\citenamefont{Li, Chernikov, Zhang,
  Rigosi, Hill, van~der Zande, Chenet, Shih, Hone, and Heinz}}]{heinz2014}
\bibinfo{author}{\bibfnamefont{Y.}~\bibnamefont{Li}},
  \bibinfo{author}{\bibfnamefont{A.}~\bibnamefont{Chernikov}},
  \bibinfo{author}{\bibfnamefont{X.}~\bibnamefont{Zhang}},
  \bibinfo{author}{\bibfnamefont{A.}~\bibnamefont{Rigosi}},
  \bibinfo{author}{\bibfnamefont{H.~M.} \bibnamefont{Hill}},
  \bibinfo{author}{\bibfnamefont{A.~M.} \bibnamefont{van~der Zande}},
  \bibinfo{author}{\bibfnamefont{D.~A.} \bibnamefont{Chenet}},
  \bibinfo{author}{\bibfnamefont{E.-M.} \bibnamefont{Shih}},
  \bibinfo{author}{\bibfnamefont{J.}~\bibnamefont{Hone}}, \bibnamefont{and}
  \bibinfo{author}{\bibfnamefont{T.~F.} \bibnamefont{Heinz}},
  \bibinfo{journal}{Physical Review B} \textbf{\bibinfo{volume}{90}},
  \bibinfo{pages}{205422} (\bibinfo{year}{2014}).

\bibitem[{\citenamefont{Laturia et~al.}(2018)\citenamefont{Laturia, Van~de Put,
  and Vandenberghe}}]{laturia2018}
\bibinfo{author}{\bibfnamefont{A.}~\bibnamefont{Laturia}},
  \bibinfo{author}{\bibfnamefont{M.~L.} \bibnamefont{Van~de Put}},
  \bibnamefont{and} \bibinfo{author}{\bibfnamefont{W.~G.}
  \bibnamefont{Vandenberghe}}, \bibinfo{journal}{npj 2D Materials and
  Applications} \textbf{\bibinfo{volume}{2}}, \bibinfo{pages}{6}
  (\bibinfo{year}{2018}).

\bibitem[{\citenamefont{Zhao and Zhang}(2017)}]{bozhao2017}
\bibinfo{author}{\bibfnamefont{B.}~\bibnamefont{Zhao}} \bibnamefont{and}
  \bibinfo{author}{\bibfnamefont{Z.~M.} \bibnamefont{Zhang}},
  \bibinfo{journal}{International Journal of Heat and Mass Transfer}
  \textbf{\bibinfo{volume}{106}}, \bibinfo{pages}{1025} (\bibinfo{year}{2017}).

\bibitem[{\citenamefont{Chernikov et~al.}(2015)\citenamefont{Chernikov,
  Ruppert, Hill, Rigosi, and Heinz}}]{chernikov2015}
\bibinfo{author}{\bibfnamefont{A.}~\bibnamefont{Chernikov}},
  \bibinfo{author}{\bibfnamefont{C.}~\bibnamefont{Ruppert}},
  \bibinfo{author}{\bibfnamefont{H.~M.} \bibnamefont{Hill}},
  \bibinfo{author}{\bibfnamefont{A.~F.} \bibnamefont{Rigosi}},
  \bibnamefont{and} \bibinfo{author}{\bibfnamefont{T.~F.} \bibnamefont{Heinz}},
  \bibinfo{journal}{Nature Photonics} \textbf{\bibinfo{volume}{9}},
  \bibinfo{pages}{466} (\bibinfo{year}{2015}).

\bibitem[{\citenamefont{Agranovich and Kravtsov}(1985)}]{agranovich1985notes}
\bibinfo{author}{\bibfnamefont{V.}~\bibnamefont{Agranovich}} \bibnamefont{and}
  \bibinfo{author}{\bibfnamefont{V.}~\bibnamefont{Kravtsov}},
  \bibinfo{journal}{Solid State Communications} \textbf{\bibinfo{volume}{55}},
  \bibinfo{pages}{85} (\bibinfo{year}{1985}).

\bibitem[{\citenamefont{Smith et~al.}(2005)\citenamefont{Smith, Vier, Koschny,
  and Soukoulis}}]{smith2005}
\bibinfo{author}{\bibfnamefont{D.}~\bibnamefont{Smith}},
  \bibinfo{author}{\bibfnamefont{D.}~\bibnamefont{Vier}},
  \bibinfo{author}{\bibfnamefont{T.}~\bibnamefont{Koschny}}, \bibnamefont{and}
  \bibinfo{author}{\bibfnamefont{C.}~\bibnamefont{Soukoulis}},
  \bibinfo{journal}{Physical Review E} \textbf{\bibinfo{volume}{71}},
  \bibinfo{pages}{036617} (\bibinfo{year}{2005}).

\bibitem[{\citenamefont{Tran et~al.}(2014)\citenamefont{Tran, Soklaski, Liang,
  and Yang}}]{BP1}
\bibinfo{author}{\bibfnamefont{V.}~\bibnamefont{Tran}},
  \bibinfo{author}{\bibfnamefont{R.}~\bibnamefont{Soklaski}},
  \bibinfo{author}{\bibfnamefont{Y.}~\bibnamefont{Liang}}, \bibnamefont{and}
  \bibinfo{author}{\bibfnamefont{L.}~\bibnamefont{Yang}},
  \bibinfo{journal}{Physical Review B} \textbf{\bibinfo{volume}{89}},
  \bibinfo{pages}{235319} (\bibinfo{year}{2014}).

\bibitem[{\citenamefont{Xia et~al.}(2014)\citenamefont{Xia, Wang, and
  Jia}}]{BP2}
\bibinfo{author}{\bibfnamefont{F.}~\bibnamefont{Xia}},
  \bibinfo{author}{\bibfnamefont{H.}~\bibnamefont{Wang}}, \bibnamefont{and}
  \bibinfo{author}{\bibfnamefont{Y.}~\bibnamefont{Jia}},
  \bibinfo{journal}{Nature Communications} \textbf{\bibinfo{volume}{5}},
  \bibinfo{pages}{4458} (\bibinfo{year}{2014}).

\bibitem[{\citenamefont{Zhang et~al.}(2016)\citenamefont{Zhang, Guo, Huang,
  Zhu, Cai, Xie, Zhou, and Zeng}}]{SiP}
\bibinfo{author}{\bibfnamefont{S.}~\bibnamefont{Zhang}},
  \bibinfo{author}{\bibfnamefont{S.}~\bibnamefont{Guo}},
  \bibinfo{author}{\bibfnamefont{Y.}~\bibnamefont{Huang}},
  \bibinfo{author}{\bibfnamefont{Z.}~\bibnamefont{Zhu}},
  \bibinfo{author}{\bibfnamefont{B.}~\bibnamefont{Cai}},
  \bibinfo{author}{\bibfnamefont{M.}~\bibnamefont{Xie}},
  \bibinfo{author}{\bibfnamefont{W.}~\bibnamefont{Zhou}}, \bibnamefont{and}
  \bibinfo{author}{\bibfnamefont{H.}~\bibnamefont{Zeng}}, \bibinfo{journal}{2D
  Materials} \textbf{\bibinfo{volume}{4}}, \bibinfo{pages}{015030}
  (\bibinfo{year}{2016}).

\bibitem[{\citenamefont{Li et~al.}(2018{\natexlab{a}})\citenamefont{Li, Wang,
  Gong, Zhu, Deng, Shi, Gao, Li, and Zhai}}]{GeP}
\bibinfo{author}{\bibfnamefont{L.}~\bibnamefont{Li}},
  \bibinfo{author}{\bibfnamefont{W.}~\bibnamefont{Wang}},
  \bibinfo{author}{\bibfnamefont{P.}~\bibnamefont{Gong}},
  \bibinfo{author}{\bibfnamefont{X.}~\bibnamefont{Zhu}},
  \bibinfo{author}{\bibfnamefont{B.}~\bibnamefont{Deng}},
  \bibinfo{author}{\bibfnamefont{X.}~\bibnamefont{Shi}},
  \bibinfo{author}{\bibfnamefont{G.}~\bibnamefont{Gao}},
  \bibinfo{author}{\bibfnamefont{H.}~\bibnamefont{Li}}, \bibnamefont{and}
  \bibinfo{author}{\bibfnamefont{T.}~\bibnamefont{Zhai}},
  \bibinfo{journal}{Advanced Materials} \textbf{\bibinfo{volume}{30}},
  \bibinfo{pages}{1706771} (\bibinfo{year}{2018}{\natexlab{a}}).

\bibitem[{\citenamefont{Li et~al.}(2018{\natexlab{b}})\citenamefont{Li, Gong,
  Sheng, Wang, Wang, Zhu, Shi, Wang, Han, Yang et~al.}}]{GeAs2}
\bibinfo{author}{\bibfnamefont{L.}~\bibnamefont{Li}},
  \bibinfo{author}{\bibfnamefont{P.}~\bibnamefont{Gong}},
  \bibinfo{author}{\bibfnamefont{D.}~\bibnamefont{Sheng}},
  \bibinfo{author}{\bibfnamefont{S.}~\bibnamefont{Wang}},
  \bibinfo{author}{\bibfnamefont{W.}~\bibnamefont{Wang}},
  \bibinfo{author}{\bibfnamefont{X.}~\bibnamefont{Zhu}},
  \bibinfo{author}{\bibfnamefont{X.}~\bibnamefont{Shi}},
  \bibinfo{author}{\bibfnamefont{F.}~\bibnamefont{Wang}},
  \bibinfo{author}{\bibfnamefont{W.}~\bibnamefont{Han}},
  \bibinfo{author}{\bibfnamefont{S.}~\bibnamefont{Yang}}, \bibnamefont{et~al.},
  \bibinfo{journal}{Advanced Materials} \textbf{\bibinfo{volume}{30}},
  \bibinfo{pages}{1804541} (\bibinfo{year}{2018}{\natexlab{b}}).

\bibitem[{\citenamefont{Yu and Fan}(2009)}]{zongfuTimeModulation2009}
\bibinfo{author}{\bibfnamefont{Z.}~\bibnamefont{Yu}} \bibnamefont{and}
  \bibinfo{author}{\bibfnamefont{S.}~\bibnamefont{Fan}},
  \bibinfo{journal}{Nature Photonics} \textbf{\bibinfo{volume}{3}},
  \bibinfo{pages}{91} (\bibinfo{year}{2009}).

\bibitem[{\citenamefont{Fang et~al.}(2012)\citenamefont{Fang, Yu, and
  Fan}}]{kejieTimeModulation2012}
\bibinfo{author}{\bibfnamefont{K.}~\bibnamefont{Fang}},
  \bibinfo{author}{\bibfnamefont{Z.}~\bibnamefont{Yu}}, \bibnamefont{and}
  \bibinfo{author}{\bibfnamefont{S.}~\bibnamefont{Fan}},
  \bibinfo{journal}{Physical Review Letters} \textbf{\bibinfo{volume}{108}},
  \bibinfo{pages}{153901} (\bibinfo{year}{2012}).

\bibitem[{\citenamefont{Sounas and Al{\`u}}(2017)}]{aluTimeModulation2017}
\bibinfo{author}{\bibfnamefont{D.~L.} \bibnamefont{Sounas}} \bibnamefont{and}
  \bibinfo{author}{\bibfnamefont{A.}~\bibnamefont{Al{\`u}}},
  \bibinfo{journal}{Nature Photonics} \textbf{\bibinfo{volume}{11}},
  \bibinfo{pages}{774} (\bibinfo{year}{2017}).

\end{thebibliography}

\end{document}